\documentclass[aps,pre,reprint,superscriptaddress,amsmath,showpacs]{revtex4-1}
\usepackage[hyperindex,breaklinks,colorlinks=true]{hyperref}
\usepackage{graphicx,sidecap,enumerate,latexsym,amsfonts,amssymb,amsmath}
\usepackage{xcolor}
\usepackage{verbatim}
\usepackage{chngcntr}

\begin{document}
\title{Dynamic topologies of activity-driven temporal networks with memory}

\author{Hyewon Kim}
\affiliation {Department of Physics,
Korea Advanced Institute of Science and Technology, Daejeon
34141, Korea}

\author{Meesoon Ha}
\email[]{msha@chosun.ac.kr}
\affiliation{Department of Physics Education, Chosun University,
Gwangju 61452, Korea}

\author{Hawoong Jeong}
\email[]{hjeong@kaist.edu}
\affiliation {Department of Physics,
Korea Advanced Institute of Science and Technology, Daejeon
34141, Korea}
\affiliation{Institute for the BioCentury, Korea Advanced Institute of Science and Technology,
Daejeon 34141, Korea}

\date{\today}

\begin{abstract}
We propose dynamic scaling in temporal networks with heterogeneous activities and memory, and provide a comprehensive picture for the dynamic topologies of such networks, in terms of the modified activity-driven network model [H. Kim \textit{et al.}, Eur. Phys. J. B {\bf 88}, 315 (2015)]. Particularly, we focus on the interplay of the time resolution and memory in dynamic topologies. 
Through the random walk (RW) process, we investigate diffusion properties and topological changes as the time resolution increases. Our results with memory are compared to those of the memoryless case. Based on the temporal percolation concept, we derive scaling exponents in the dynamics of the largest cluster and the coverage of the RW process in time-varying networks. We find that the time resolution in the time-accumulated network determines the effective size of the network, while memory affects relevant scaling properties at the crossover from the dynamic regime to the static one. The origin of memory-dependent scaling behaviors is the dynamics of the largest cluster, which depends on temporal degree distributions. Finally, we conjecture of the extended finite-size scaling ansatz for dynamic topologies and the fundamental property of temporal networks, which are numerically confirmed. 
\end{abstract}

  
        
  
\maketitle

\section{Introduction}
\label{intro}

Network model studies are useful for understanding complex systems, consisting of interacting components. Most studies have been done on static networks where the topology is fixed in time. Such studies have provided a wealth of information for time-invariant topological properties of the systems~\cite{Albert2002,*Barabasi2008,*Newman2010}. However, complex systems often change in time.  As technology advances, it has become possible to collect time-series data. Recent studies have started discussing on the characteristics of time-varying (temporal) networks~\cite{Iribarren2009,*Cattuto2010, *Rocha2010, *Stehle2011, *Karsai2011, *Jo2012, *Pfitzner2013, *Starnini2013a, *Scholtes2014, *Valdano2015, *Saramaki2015}. Temporal network studies have demonstrated both the importance of the temporal dimension and the necessity for studying  physically relevant properties~\cite{Holme2012, *Holme2015} on demand. 

Temporal networks are expressed as the time series of network formations, where the embedded topological and temporal information can be so different from one to another. Since the topology is one of the important features to understand network properties, it is very meaningful to study the dynamic topologies of temporal networks as well as the time resolution that plays a crucial role in analyzing dynamic topologies~\cite{Salathe2010,* Rasmussen2013,*Masuda2013, *Chen2015, *Masuda2017}.

One way to study the interplay of the time resolution and network topology is to set the size of time window $t_w$, within which the detailed temporal information is ignored~\cite{Caceres2011, *Krings2012, *Holme2013, Ribeiro2013}. For an example, if $t_w$ is large enough to contain all the interactions in a single snapshot, the temporal network can be considered as a time-aggregated (static) network, where network dynamics is neglected. If $t_w$ is small, networks change in a short time and the dynamic properties of the networks depend on $t_w$. 
To figure out the proper time resolution $t_w$ that well describes the dynamic topologies of temporal networks, it is often considered the random-walk (RW) process. It is because the RW process is one of the fundamental tools to analyze the diffusion property of underlying structures. Thus the RW process has been widely discussed on static~\cite{Almaas2003, *Baronchelli2008, Bacco2015} and temporal networks~\cite{Starnini2012, Ribeiro2013, *Perra2012a, *Delvenne2015, *Alessandretti2017}. 
However, dynamic scaling for diffusion properties in temporal networks is more complicated than that in static ones, which is attributed to topological changes. The path of the walk depends on network structures, so that physically relevant quantities, such as the mean first passage time and the coverage (the number of distinct visited nodes by the walk), reveal the origin of dynamic scaling. Hence it is an important task to speculate dynamic scaling in temporal networks according to $t_w$.

In many interacting systems, it has been reported that a community structure can be developed according to individual activities and accumulated experiences (kind of memory)~\cite{Onnela2007, *Sun2013, *Vestergaard2014}. To mimics the time evolution of such a structure, temporal network models have been suggested with and without memory, distinct from static ones~\cite{Perra2012, Moinet2015, *Gleeson2016, Karsai2014, *Medus2014, *Laurent2015, Ubaldi2016, Kim2015}. Among them, the activity-driven network (ADN) model~\cite{Perra2012} is first introduced as the simplified version that generates the structural heterogeneity of highly dynamical networks by characterizing interaction patterns of nodes.
Later, it has been modified with the consideration of various memory-type parameters to explain the detailed parts in non-Markovian empirical data of real networks~\cite{Karsai2014, *Medus2014, *Laurent2015, Ubaldi2016, Kim2015}.  

\begin{figure}[]
\center
\includegraphics[width=\columnwidth]{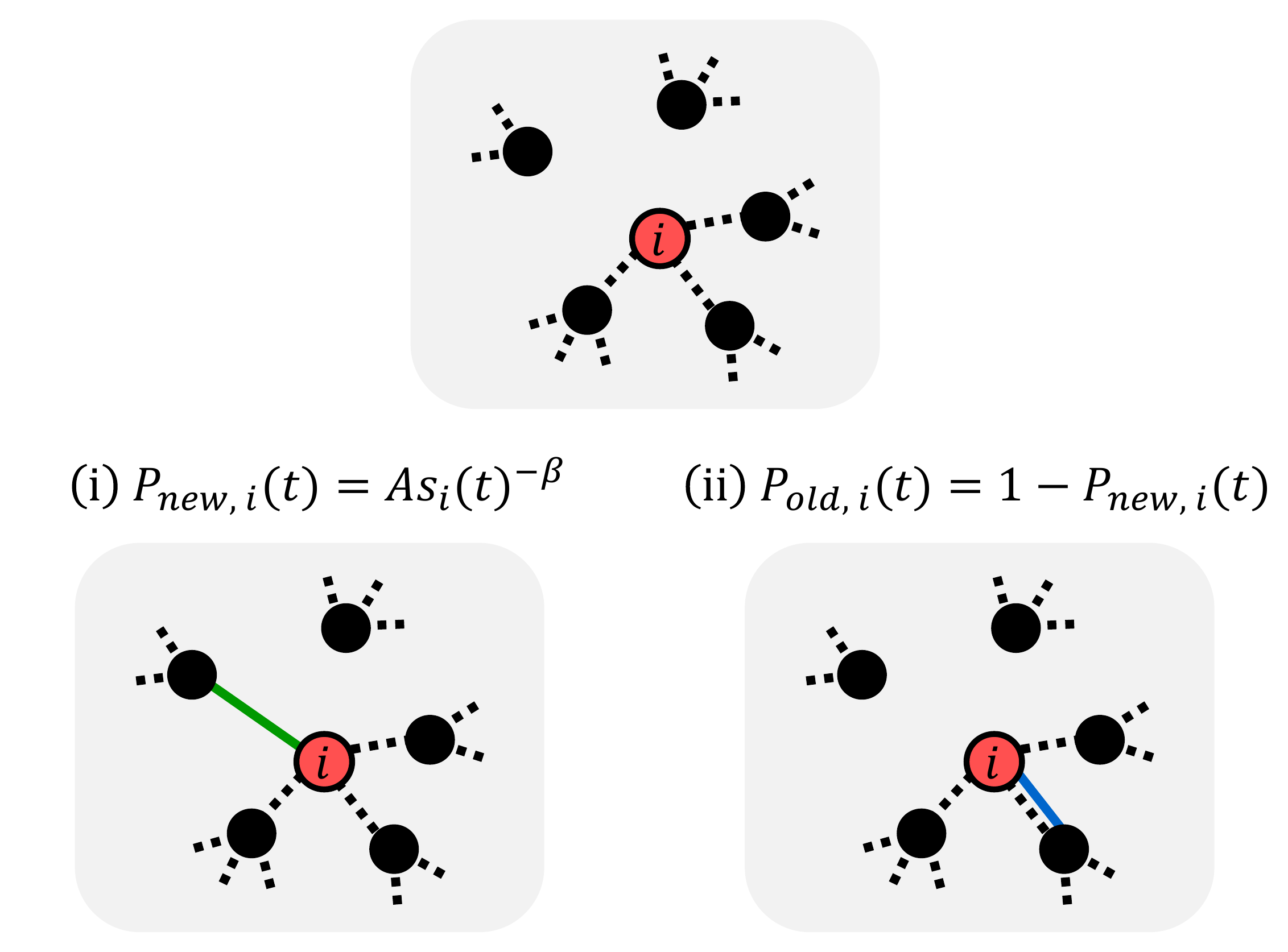}
\caption{The modified ADN model is schematically illustrated. If the node $i$ becomes active with $s_i(t)$ that is the accumulated strength of the node $i$ up to time $t$ in the upper panel, it can choose one of two possible ways to make a link with the following probability:  
(i) $P_{\rm new}(t)=A s_i(t)^{-\beta}$ and (ii) $P_{\rm old}(t)=1-P_{\rm new}(t)$, where $A$ is the model coefficient and $\beta$ is the memory exponent. The active node is highlighted as a different colored (red) symbol. While the new (green) link is created at random, the old (blue) link is created with the preferential probability. Dashed lines represent old links created before time $t$.}
\label{fig1}
\end{figure}

In this paper, we revisit the modified ADN model (see Fig.~\ref{fig1}) to discuss the origin of scaling properties in temporal networks with memory. The modified ADN model in our early study~\cite{Kim2015} is generated by linking memory as well as nodal activities. The main result of the model was compared to the detailed structural properties of empirical data. Moreover, it was reported that there are scaling properties as the time resolution varies. However, our previous finding was more or less preliminary since one specific case (relatively weak memory) was tested without the finite-size effect. Hence the validity check is necessary as well as either the analytic conjecture or the physical argument of scaling properties according to different memory exponents.

Particularly, we here focus on the interplay of the time resolution and memory in dynamic scaling. It is found that the time resolution determines the crossover of relevant scaling properties, while memory directly affects the growth of the largest cluster due to network structural changes. The shorter the time resolution is, the more dynamic the topological change occurs, in the context of the effective size of the network. As memory gets stronger, the growth of the largest cluster gets slower because the largest cluster is affected by the structural change of temporal degree distributions. 

Our results are compared with those in temporal percolation~\cite{Starnini2014}, where the percolation properties were studied in the ADN model with heterogeneous nodal activities as well as the relevance of degree correlations. Based on the expression of percolation time marking the onset of the giant connected component (GCC), we also discuss the temporal percolation threshold and degree correlations. In the vicinity of the rescaled percolation threshold, the finite-size scaling (FSS) of the GCC is governed by the FSS exponent of the critical shift from the thermodynamic threshold. As a result, we are able to conjecture the values of scaling exponents by the power-counting analysis from the complete form of FSS, and propose the fundamental property of dynamic scaling, which are numerically confirmed.

The rest of this paper is organized as follows. In Sec.~\ref{model}, we describe the modified ADN model and discuss the role of memory in network structural properties. In Sec.~\ref{diffusion}, we show how to define subnetworks according to $t_w$, consider the RW process and temporal percolation on such networks, and conjecture dynamic scaling for them. In Sec.~\ref{numerics}, we present numerical results.  In Sec.~\ref{conjecture}, the fundamental property is proposed to provide a comprehensive picture for dynamic topologies of temporal networks. Finally, we conclude this paper with summary and some remarks in Sec.~\ref{summary}. Appendix~\ref{appendix} provides mathematical derivations of network properties with extra plots for the time-aggregated version of the modified ADN model. 

\section{Temporal Network Model}
\label{model}

Consider the modified ADN model with the power-law distributions of nodal activities, which presents the emergence of weighted scale-free networks in the time-aggregated picture~\cite{Kim2015}.  A temporal network of $N$ disconnected nodes is generated with the following dynamics:  Each node $(i\in \{1, ...,N\})$ is assigned to its own activity with an active rate, $a$, which follows an activity distribution, $F(a)\sim a^{-\gamma}$ with $\gamma> 2$. In addition, the node is described by its degree $k$, strength $s$, and link-weight $w$.
At each step, each node $i$ activates with the probability $p_i=a_i/a_M$, where $a_M=\max_{j=1}^{N}\{a_j\}$. Once the node $i$ becomes active, it creates a link with other nodes except itself. If the node $i$ activates for the first time, it connects to an arbitrary node at random. Otherwise, it creates a link by either (i) or (ii) (see Fig.~\ref{fig1}): 
\begin{enumerate}[(i)]
\item{ {\bf New link} -- With probability $P_{{\rm new},i}(t)$, the active node $i$ connects to a node at random among the nodes that have never connected to the node $i$ in the past time up to time $t$.}
\item{ {\bf Old link} -- With probability $P_{{\rm old},i}(t)=1-P_{{\rm new},i}(t)$, the active node $i$ connects to one of ever connected nodes, say the node $j$, with probability $\Pi_{ij}$ that is proportional to the frequency of the links between the node $i$ and the node $j$ up to time $t$.}
\end{enumerate}
After this step, the connected nodes and links are formed as a network at time $t$. If two nodes $i$ and $j$ are connected at time $t$, $\mathcal{A}_{ij}(t)=1$, otherwise $\mathcal{A}_{ij}(t)=0$. The nodal memory, the information of the generated links, $\mathcal{A}_{ij}(t)$, is updated. Once all information for $N$ nodes is recorded, the generated links are deleted and time is updated as $t+1$. The above procedure is repeated until $t=T$. 

Based on the empirical dataset analysis~\cite{Kim2015}, we set  
\begin{align}
P_{{\rm new},i}(t)=As_i(t)^{-\beta}, 
\label{pnew}
\end{align}
where $A$ is the model coefficient (in this work, $A=1$ from now on without loss of generality), $s_i(t)$ is the accumulated strength of the node $i$ up to time $t$, \textit{i.e.}, $s_i(t)= \sum_{t'=0}^{t}\sum_{j=1}^{N}\mathcal{A}_{ij}(t')$, and $\beta$ is the memory exponent with $0\leq \beta \leq1$. The preferential probability, $\Pi_{i,j}(t)$, is defined as follows: 
\begin{align}
\Pi_{ij}(t)=\frac{w_{ij}(t)}{\sum_{l\in\mathcal{N}_i(t)}w_{il}(t)}, 
\label{pold}
\end{align}
where $\mathcal{N}_i(t)$ is a set for the nodes that have ever been connected to the node $i$ up to time $t$ and $w_{ij}(t)$ is the accumulated link-weight between the node $i$ and $j$ up to time $t$, \textit{i.e.}, $w_{ij}(t)=\sum_{t'=0}^{t}\mathcal{A}_{ij}(t')$. 

Equation~\eqref{pnew} represents how much the node $i$ prefers to make a new friend at time $t$. If $\beta=0$, then $P_{{\rm new},i}(t)$ becomes a time-independent constant, so that one can be called the memoryless case. On the other hand, for $\beta \neq 0$, the node $i$ prefers to meet old friends, rather than a new one, and Eq.~\eqref{pold} indicates that the node $i$ has a different preference among its friends. To sum up, there are two main control parameters,
$\gamma$ and $\beta$, associated with the static and dynamic characteristics of temporal networks.
The activity exponent $\gamma$ is related to the heterogeneity of nodal activities, and the memory exponent $\beta$ controls the strength of inter-personal social ties in communities. 

The generated temporal network can yield the power-law distributions of strength ($s$), degree ($k$), and link-weight ($w$) in the time-aggregated picture with decay exponents 
$\gamma_s$, $\gamma_k$, and $\gamma_w$, respectively: $P(s) \sim s^{-\gamma_s}$ with $\gamma_s=\gamma$, $P(k) \sim k^{-\gamma_k}$ with $\gamma_k=\frac{(\gamma-\beta)}{(1-\beta)}$ and $P(w) \sim w^{-\gamma_w}$ with $\gamma_w=\gamma$ as $\beta \rightarrow 1$ (see detailed mathematical derivations in Appendix~\ref{appendix}).

Network structures change in time and the amount of such structural changes depends on the time resolution. To understand the impact of the time resolution on dynamic topologies, we control it as the time-window size $t_w$ in temporal networks, and perform the RW process that is one of the useful tools to capture the structural and temporal properties in networks. In addition, we consider two types of networks without and with memory to speculate the role of memory in dynamic scaling.
The range of the memory exponent is $0\leq \beta \leq1$, but we only consider up to $\beta= 0.5$ for the memory case. It is because the case of $\beta=1$ yields the exponential degree distribution rather than the power-law one, which is out of our interest. Moreover, we are interested in $2\le \gamma\le 3$, which is the most heterogeneous case with the finite average of activities in the thermodynamic limit.
Without loss of generality, we choose $(\gamma, \beta)$ as (2.5, 0) for the network without memory and (2.5, 0.5) for the case of the strongest memory that yields the power-law degree distribution $P(k)\sim k^{-\gamma_k}$: $P(k)\sim k^{-2.5}$ for $\beta=0$ and $P(k)\sim k^{-4.0}$ for $\beta=0.5$.

\section{Diffusion and Scaling Properties}
\label{diffusion}

A temporal network is the combination of time-ordered subnetworks which are determined by $t_w$. As $t_w$ increases, the temporal information of the network disappears. If $t_w$ is large enough to capture all the interactions into a single subnetwork, then the temporal network can be treated as a static network. So by adjusting $t_w$, we can reconstruct the temporal network from highly dynamic network to static one.

\begin{figure}[b]
\center
\includegraphics[width=\columnwidth]{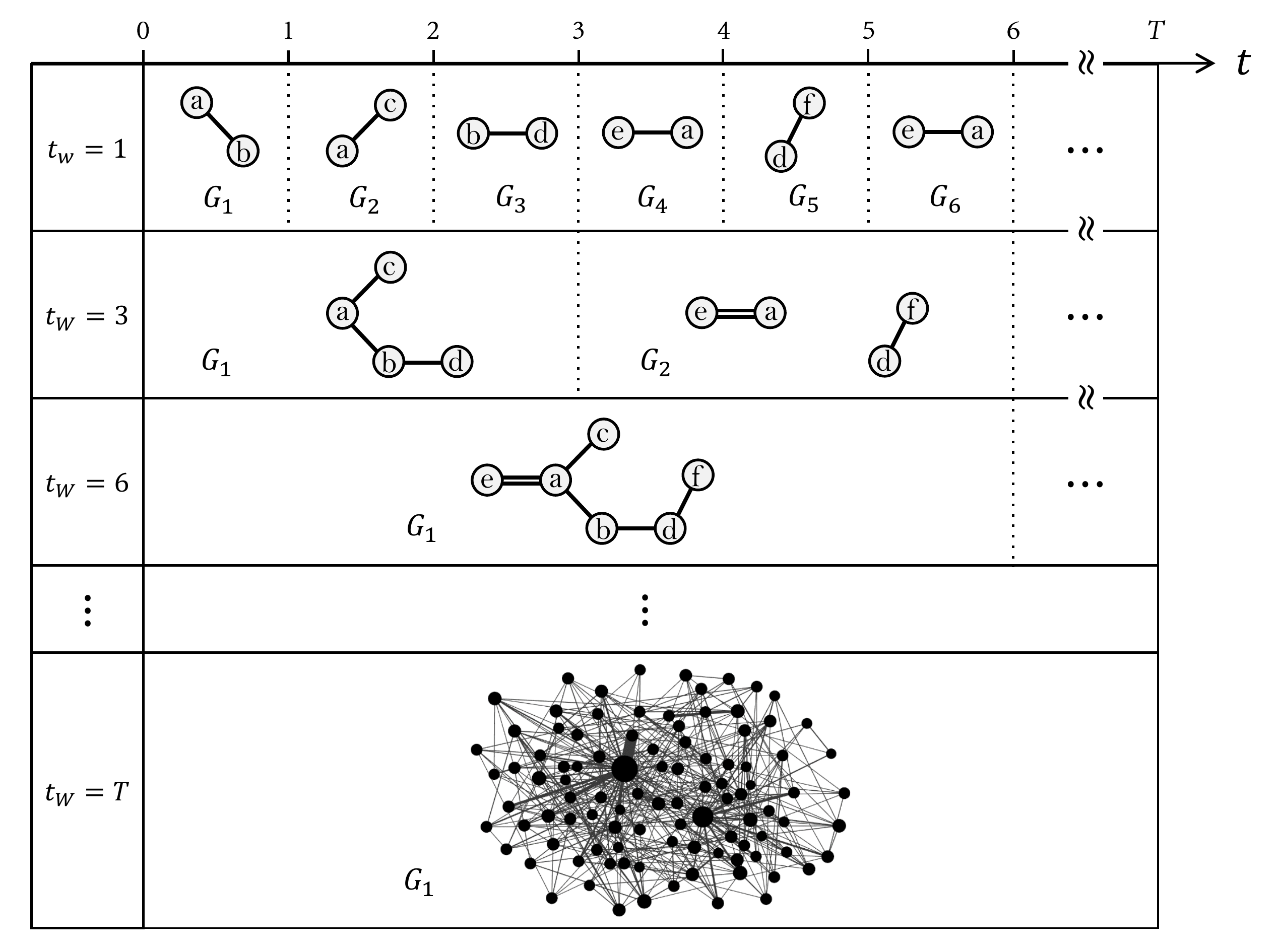}
\caption{The schematic illustrations of temporal networks are shown for various $t_w$ ($1\le t_w\le T$). The time resolution $t_w$ controls the dynamic level of the networks. As $t_w$ increases, the topological information is developed as a static property and the temporal information is negligible. When $t_w$ is large enough to contain all the interactions in a single snapshot, say $t_w=T$, it corresponds a static network, where the detailed temporal information can be ignored. The thickness of the link represents the link-weight value. At $t_w=T$, the topology of the modified ADN model is illustrated for $N=100$ and $T=3000$ with $\gamma=2.5$ and $\beta=0.5$.}
\label{fig2}
\end{figure}

In Fig.~\ref{fig2}, we schematically illustrate coarse-grained temporal networks for various $t_w$. Once $t_w$ is set, the temporal network is represented as a series of the subnetwork $G_n (n=1, 2, ..., \lceil T/t_w \rceil)$, which is the accumulated network during the interval $[(n-1)t_w, nt_w)$. Here the strength and the degree of the node $i$ in $G_n$ are denoted by $s_i(G_n)$ and $k_i(G_n)$, respectively, and $w_{ij}(G_n)$ represents the link-weight between node $i$ and $j$ in $G_n$. 
 
For coarse-grained networks, we study the effect of the time resolution on dynamic processes. Particulary, we are interested in the influence of memory according to the time resolution. Hence we define the number of events (links) as time $t$ to facilitate the comparison of the results of the network without memory to those with memory. It allows that all the subnetworks for the same $t_w$ have the same number of links, and we can directly compare their interaction patterns.

\subsection*{Random-Walk (RW) Process}

To speculate diffusion in temporal networks as $t_w$ varies, we perform the RW process on coarse-grained temporal networks within  $t_w$. A walker initially starts at a randomly selected node, and moves on subnetworks of the temporal network determined by $ t_w $. The walker moves $t_w$ steps on each subnetwork, $G_n$. The transition probability from the current node to the target node is proportional to the link-weight in $G_n$. We measure the number of distinct visited nodes (the coverage) by the walker, $V(t,N)$ at time $t$ for the network of $N$ nodes. 

The dynamics of the walker is as follows: Consider a subnetwork, $G_n$, that is an accumulated network during the time interval $[(n-1)t_w, nt_w)$, and assume that the walker is located the node $i$. If the node $i$ has links in $G_n$, \textit{i.e.}, $s_i(G_n)\neq0$, then the walker moves from the node $i$ to one of its neighboring nodes, say $j$, with probability 
$p_{ij}(G_n)=w_{ij}(G_n)/s_i(G_n).$
Otherwise, it remains on the node $i$ until its link is created. After $t_w$ steps, the subnetwork $G_n$ is switched to the next subnetwork $G_{n+1}$ and the position of walker is updated by the same procedure. We repeat the dynamics until the last subnetwork, $G_{\lceil T/t_w \rceil}$. Here the total number of links is $T$ (the maximal time, sufficiently large).

The coverage $V$ of the walker is accumulated as counting the distinct nodes that the walk has ever visited as time elapses, which follows the extended FSS ansatz:
\begin{align}
\label{FSS}
V(t,N)=N^{\alpha}f(\tau),
\end{align} 
where $\tau\equiv t/N$.

Assuming that there is the crossover time of dynamic topologies from the dynamic regime to the static regime at $t_\times$, one can easily categorize temporal regimes: In the static regime $(t_w\gg t_\times)$ with $\alpha=1$, $f(\tau)\sim \tau$ for $\tau\ll 1$ and $f(\tau)=\text{constant}$ for $\tau\gg 1$, whereas in the dynamic regime $(t_w\ll t_\times)$ with $0\le\alpha<1$, $f(\tau)\sim \tau^{\zeta}$ with $0\le \zeta<1$ for $t<t_w$ and $\zeta=1$ for $t>t_w$. The values of $\zeta$ and $\alpha$ depend on $t_w$ and $\beta$. This conjecture is based on 
the preliminary test with weak memory ($\beta=0.12$)~\cite{Kim2015} and the typical scaling of static networks~\cite{Almaas2003, Baronchelli2008}, where $V\sim N\tau$ for $\tau\ll 1$, the finite-size effect comes around $\tau\approx 1$, and finally $V\to N$ for $\tau\gg 1$. To understand the anomalous scaling behaviors of $V$ in the intermediate time regime $(t<t_w<t_\times)$, we analyze dynamic topologies as $t_w$ varies, in the context of network and temporal percolation properties.

\subsection*{Network Structural Properties}

To analyze the interplay of the time resolution and memory in the RW process, it is necessary to discuss the structural change of the accumulated network up to $t_w$, in terms of the temporal degree distribution, the average degree, and the number of distinct nodes, denoted as $P_w(k), \langle k(t_w)\rangle,$ and $N_d(t_w)$, respectively.

\subsubsection*{Temporal Degree Distribution}

The dynamic topologies of temporal networks can be described as temporal degree distributions, which satisfy the following scaling forms:
\begin{align}
\label{Pw}
P_w(k;N)=
\left\{
\begin{array}{lr}
t_w^{-1}\phi_{-}(k/k_c)\\
t_w^{-\theta}\phi_{+}(k/k_p)
\end{array}
\begin{array}{ll}
& \text{for}~t_w\ll t_\times,\\
& \text{for}~t_w\gg t_\times,
\end{array}
\right.
\end{align}
where $t_\times$ is the crossover time of the network growth ($t_\times\sim N^{z}$), $k_c$ is the strong cutoff of degrees ($k_c\sim t_w^{1/\gamma_k}$), and $k_p$ is the most probable degree ($k_p\sim t_w^{\theta}$). The scaling functions should satisfy $\phi_{\mp}(\kappa) \sim {\kappa}^{-\gamma_k}$ with $\gamma_k=\frac{\gamma-\beta}{1-\beta}$ from mathematical derivations in Eq.~\eqref{PsPk}, 
while $\theta=1-\beta$ from our previous conjecture~\cite{Kim2015}. 

\subsubsection*{Average Degree}

From temporal degree distributions, the average degree $\langle k(t_w;N)\rangle$ can be estimated. However, it would be better to consider the proper average that is rescaled by the number of distinct nodes appeared up to $t_w$, $N_d(t_w;N)$, instead of $N$:

\begin{align}
\label{Nd}
N_d(t_w;N) \simeq \left\{
\begin{array}{lr}
t_w\\
N
\end{array}
\begin{array}{ll}
& \textrm{for}~t_w\ll t_\times, \\
& \textrm{for}~t_w\gg t_\times.
\end{array}
\right.
\end{align}
Therefore,
\begin{align}
\label{kw}
\langle k(t_w)\rangle_d
\sim \left\{
\begin{array}{lr}
\text{constant}\\
t_w^{\theta}
\end{array}
\begin{array}{ll}
& \textrm{for}~t_w\ll t_\times, \\
& \textrm{for}~t_w\gg t_\times.
\end{array}
\right.
\end{align}
Here the rescaled average degree is defined as
$$\langle k(t_w)\rangle _d \equiv \langle k(t_w) \rangle N/N_d(t_w;N)$$ with 
$$\langle k \rangle\equiv \frac{1}{N}\sum_{k} kP(k,t_w;N).$$  
Equation~\eqref{kw} indicates that the memory effect on the average degree is negligible for $t_w\ll t_\times$ if it is properly rescaled by the effective network size. However, the heterogeneity of degree distributions is governed by the memory exponent, which is shown in Eq.~\eqref{Pw}. It means that even if the same number of links is added, the network growth is affected by the memory exponent.  

To reveal the underlying relevant scaling of dynamic topologies and explain the diffusion properties of the RW process,  
we take the temporal percolation concept~\cite{Starnini2014} and estimate the growth pattern of the largest cluster size in the modified ADN model for three temporal regimes and two limiting cases of memory ($\beta=0$ and $\beta=0.5$). Hence the diffusion properties of the RW process rely on $N_d$, within which the walker can only move onto another node with the connectivity in the cluster.

\begin{figure*}[]
\center
\includegraphics[width=\textwidth]{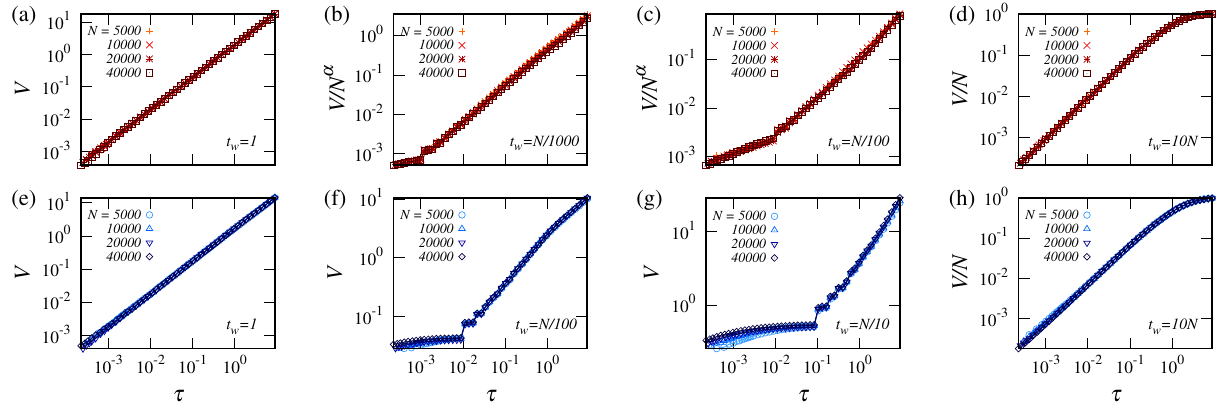}
\caption{The FSS forms are tested in the diffusion property of the RW process on the modified ADN model 
for $\gamma=2.5$, in terms of the coverage, $V,$ and the rescaled time $\tau\equiv t/N$ for $N=\{5000, 10000, 20000, 40000\}$, as $t_w$ increases up to $T=10N$. For $\beta=0$ (memoryless), (a) $t_w=1$, (b) $t_w=N/1000$, (c) $t_w=N/100$, and (d) $t_w=10N$. For $\beta=0.5$ (strong memory), (e) $t_w=1$, (f)  $t_w=N/100$, (g) $t_w=N/10$, and (h) $t_w=10N$. It is noted that both cases exhibit the same behaviors of $V$ for $t_w=1$ (highly dynamic) and $t_w\gg N$ (static). In the intermediate regime of $ t_w$, $V$ yields crossover behaviors around $t=t_w$. Moreover, the memoryless case ($\beta=0$) shows finite-size effects with a non-trivial exponent $\alpha$, varying from (b) $\alpha=0.2$ to (c) $\alpha=0.4$. Numerical data are averaged over 100 network configurations and $2\times 10^3$ runs with different initial nodes per network.}
\label{fig3}
\end{figure*} 

\subsection*{The Largest Cluster:\\ Giant Connected Component}

The largest cluster size of the accumulated network up to $t_w$, namely the size of GCC, is denoted as $M(t_w, N)$. The dynamics of $M$ is categorized as follows: In highly dynamic regime, $M$ is proportional to the natural cutoff of the effective size of the network ($t_w$), irrespectively of $N$. Then we conjecture that $M\simeq k_{\rm nat}(t_w) \sim t_w^{1/\omega_{\rm nat}}$ with $\omega_{\rm nat}=\gamma_k-1$ in the dynamic regime, while in the static regime, it scales as $M=N\mu(\tau_w)$ 
with $\tau_w\equiv t_w/N$; $\mu(\tau_w)=\text{constant}$ for $\tau_w\gg \tau_\times$. The scaling function $\mu$ is only valid for $\tau_w\ll \tau_\times$ (before the rescaled crossover time, $\tau_\times$, from the dynamic regime to the static one). In the intermediate time regime ($ 1\ll \tau_w \ll \tau_\times$), the fraction of the GCC, $m\equiv M/N$, satisfies dynamic scaling as follows: 
\begin{align}
\label{m}
m(t_w,N)\sim N^{-y}g(\tau_w N^{x}),
\end{align}
where $g(X)\sim X^{\eta}$ with $\eta=\omega_{\rm nat}^{-1}$ for $X\ll 1$, provided that $\tau_\times \sim N^{-x}$, \textit{i.e.}, $t_\times\sim N^{1-x}$, in the dynamic regime.

Since the dynamic and static exponents $x$ and $y$ can be predicted by the temporal percolation concept, we get memory-dependent in dynamic scaling: $x=1/\bar{\nu}_p$  and $y=\beta_p/\bar{\nu}_p$, where we use the critical exponents ($\bar{\nu}_p, \beta_p$) of the percolation universality class in scale-free networks~\cite{Cohen2002}: $\bar{\nu}_p=|\frac{\omega}{3-\gamma_k}|, \beta_p=|\frac{1}{3-\gamma_k}|$ for $2<\gamma_k<5$, where $\omega$ is the cutoff exponent of the upper degree. 

Using the power-counting analysis with $x$ and $\eta$, we derive the value of $y$ separately: 
\begin{align}
\label{y}
\frac{M}{N}\sim N^{-y}(\frac{t_w}{N}N^x)^{\eta} \to
M\sim t_w^{\eta} N^{1-y+(x-1)\eta} \sim t_w^{\eta},
\end{align}
with $y=1+(x-1)\eta$ in the dynamic regime. 
For the network with $\gamma=2.5$ and $\beta=0~(0.5)$, $\gamma_k=2.5~(4)$ and $\omega=2~(4)$. Hence $\eta=2/3$, $x=1/4$~\footnote{The results of the original ADN model~\cite{Starnini2014} are different from those in the modified ADN model with memory due model details, {\textit e.g.}, such as degree correlations and memory.} and $y=1/2$ for the memoryless case ($\beta=0$), while for strong memory ($\beta=0.5$), $\eta=1/3$, $x=1/4$~\footnote{For the case of $\gamma_k=4$ (integer), the scaling might contain some logarithmic corrections.} and $y=3/4$. 

\section{Numerical Results}
\label{numerics}

In this section, our conjectures for the extended FSS of physical quantities in Sec.~\ref{diffusion} are numerically confirmed in terms of the modified ADN model. Moreover, we reveal dynamic scaling in temporal networks, and address how the diffusion properties of the RW process is explained by the growth of the GCC according to the time resolution and memory. We finally present temporal percolation for three different time regimes and two limiting cases of memory.  

\begin{figure*}[]
\center
\includegraphics[width=\textwidth]{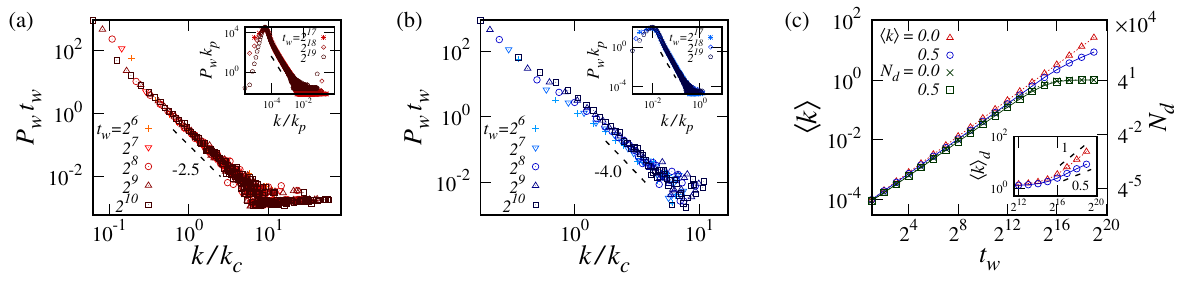}
\caption{The scaling properties of temporal degree distributions $P_w$  are tested for the wide range of $t_w$ in the modified ADN model with $\gamma=2.5$ for (a) $\beta=0$ (memoryless) and (b) $\beta=0.5$ (strong memory). In (a) and (b), the main plots represent the dynamic regime ($t_w\ll t_\times$), and the insets represent the static regime ($t_w\gg t_\times$), respectively. Dynamic scaling is based on Eq.~\eqref{Pw} with $(\gamma_k=\frac{\gamma-\beta}{1-\beta}, \theta=1-\beta)$: (a) (2.5, 1) and (b) (4, 0.5), respectively. The slopes of dashed lines correspond  the values of $\gamma_k$ for $\gamma$ with and without memory. In the main plots of (a) and (b), $k_c=t_w^{1/\gamma_k}$ (strong cutoff degree), and in the insets, $k_p=t_w^{\theta}$ (most probable degree).  (c) The average degree $\langle k\rangle$ and the total number of distinct active nodes $N_d$ are plotted as $t_w$ varies in the main plot, and the inset shows the properly rescaled average degree $\langle k\rangle_d =\langle k\rangle N/N_d(\sim k_p= t_w^{\theta})$ in the static regime, where the slopes of dashed lines correspond the values of $\theta$: 1 for $\beta=0$ and 0.5 for $\beta=0.5$, respectively. Numerical data are obtained for $N=40000$ with 500 samples.}
\label{fig4}
\end{figure*}

\subsection*{Dynamic Scaling of Coverage}

In Fig.~\ref{fig3}, the FSS form of the coverage $V(t,N)$ of the RW process against the rescaled time $\tau(\equiv t/N)$, by Eq.~\eqref{FSS}, is tested for $N\in\{5000,10,000, 20,000, 40,000\}$ for four different setting of $t_w$~\footnote{The time resolution $t_w$ is a controllable parameter, and $t$ is the accumulated number of trials in the RW process.}: For (a)--(d), $\beta=0$ at $t_w=1, N/1000, N/100,$ and $10N$ (from left to right), and for (e)--(h), $\beta=0.5$ at $t_w=1, N/100, N/10,$ and $10N$ (from left to right).  In  the highly dynamic regime ($t_w=1$), (a) and (e), $V$ exhibits trivial temporal scaling, irrespective of $N$, while in the static regime ($t_w\gg N$), (d) and (h), $t_w$ is large enough to approximate a temporal network as a static network, and $V$ shows typical static scaling as reported in the early studies~\cite{Almaas2003, Baronchelli2008}: It is known that $V\sim t$ is crossover to $V\to N$. However, in the intermediate regime ($1\ll t_w \ll N$), $V$ has a non-trivial FSS form that depends on $\beta$ and $t_w$. 

In Figs.~\ref{fig3}(b), \ref{fig3}(c), \ref{fig3}(f), and \ref{fig3}(g), we find $0\le \alpha < 1$, the value of which depends on $t_w$ and $\beta$. For the case of memoryless ($\beta=0$), we observe that $\alpha$ gradually increases as the window of the time resolution gets wider. Numerical data collapse well with $\alpha=$0.0, 0.2, 0.4 and 1 as shown in Figs.~\ref{fig3}(a)--\ref{fig3}(d), respectively. However, in the case of the network with strong memory ($\beta=0.5$), $V$ is almost unchanged up to $t_w=N/10$ with $\alpha=0$, and rapidly increases later on, as shown in Figs.~\ref{fig3}(e)--\ref{fig3}(h). 
The $\beta$-dependent anomalous scaling behaviors seem to be due to the difference in topological properties 
within the time resolution $t_w$ of temporal networks. 

To understand such anomalous behaviors of $V,$ it would be better to check the detailed network properties as $t_w$ varies with and without memory. According to $t_w$ and $\beta$, we measure the scaling properties of temporal degree distributions and other related quantities, based on the effective size of networks and the scaling form suggested in Sec.~\ref{diffusion}.

\subsection*{Scaling Behaviors of Network Properties }

In Figs.~\ref{fig4}(a) and \ref{fig4}(b), we show the scaling collapse of temporal degree distributions $P_w$ for two different time regimes by Eq.~\eqref{Pw}: The main plots in the dynamic regime and the insets in the static regime. Numerical data are measured in the modified ADN model of $N=40000$ with $\gamma=2.5$ for (a) $\beta=0$ (memoryless) and (b) $\beta=0.5$ (strong memory). Here the time resolution $t_w$ plays a crucial role in determining the effective size of the accumulated network up to $t_w$, which is so-called the number of distinct nodes, $N_d(t_w;N)$, defined in Eq.~\eqref{Nd}. 

Our numerical results show that in the dynamic regime ($t_w\ll t_{\times}$), $P_w=t_w^{-1}\phi_-(k/k_c)$ with the strong cutoff  $k_c= t_w^{1/{\gamma_k}}$, while in the static regime ($t_w\gg t_{\times}$), $P_w=t_w^{-\theta}\phi_+(k/ k_p)$ with the most probable degree $k_p=t_w^{\theta}$~\footnote{In general, the upper cutoff of degrees is denoted as $k_c\sim t_w^{1/\omega}$ with $1\le \omega\le \gamma_k$, when the network size is $t_w$.}. In Figs.~\ref{fig4}(a) and \ref{fig4}(b), it is also found that $\phi_{\pm}(\kappa_{\pm})\sim \kappa_{\pm}^{-\gamma_k}$ 
with the relative degree $\kappa_{\pm}$. 

In order to discuss the detailed statistical properties of $P_w$, we measure the average degree $\langle k\rangle$ and the number of distinct nodes $N_d$ as the function of $t_w$, which are main plots in Fig.~\ref{fig4} (c). The scaling relation of two quantities is provided by $\langle k\rangle_d$ in the inset. For $t_w\ll t_\times$, $\langle k\rangle_d\approx 1$, while $t_w\gg t_\times$, $\langle k\rangle_d \sim t^\theta$.

The topological change in the network growth can be also described by the formation of the largest cluster in percolation, namely the GCC of the network. Based on the GCC analysis, we measure temporal percolation in the modified ADN model~\cite{Kim2015}. 

\begin{figure*}[]
\center
\includegraphics[width=\textwidth]{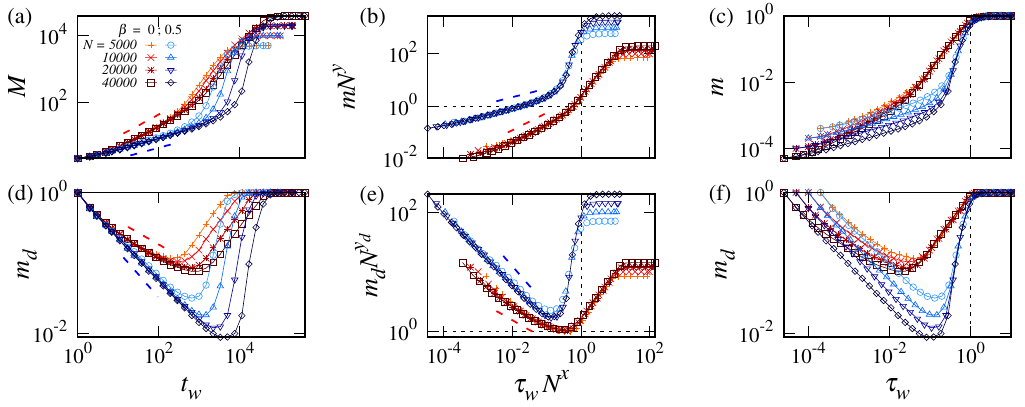}
\caption{The growth patterns of the GCC, $M$, are measured for $N\in\{5000,10,000, 20,000, 40,000\}$, where $m\equiv M/N$ and $m_d\equiv M/N_d$: For (a)--(c),  the dynamics of the GCC is plotted with conjectured FSS forms from the dynamic regime to the static regime, while for (d)--(f), the rescaled fractions by the relevant network size are shown. In the intermediate regime, we find non-trivial scaling exponents $(x,y)$ and $(x,y_d)$, respectively: (b) (1/4,~1/2) for the case of $\beta=0$ and (1/4,~3/4) for the case of $\beta=0.5$; (e) (1/4,~1/4) for $\beta=0$ and (1/4,~1/2) for $\beta=0.5$. It is noted that for $\beta=0.5$, we consider the logarithmic correction in $x=1/4$ as well. In (a)--(b) and (d)--(e), the slopes of dashed lines are correspond (2/3, 1/3) and (-1/3, -2/3) from top to bottom, respectively. Numerical data are averaged over 500 network samples.}
\label{fig5}
\end{figure*}

\subsection*{Dynamic Scaling of Percolation}

In the time-accumulated network up to $t=t_w$, the size of GCC is denoted by $M(t_w)$. The dynamics of $M$ can be categorized as three different regimes, where numerical data differently collapse in each regime: In the highly dynamic regime, $M$ only depends on $t_w$, irrespective of $N$ (the total number of nodes). In the late static regime, $m=\mu(\tau_w)$ where $\tau_w(\equiv t_w/N)$. In the inter-mediate time regime, $m=N^{-y}g(\tau_wN^{x})$ by Eq.~\eqref{m}. 

We present the scaling behaviors of $M, m(\equiv M/N),$ and $m_d(\equiv M/N_d)$ in Fig.~\ref{fig5}: In the dynamic regime, $M\sim t_w^{\eta}$ as $t_w$ increases, where the growing exponent $\eta$ depends on $\beta$: $\eta\simeq 2/3$ for $\beta=0$ and $\eta\simeq 1/3$ for $\beta=0.5$, which means that active nodes form memory-dependent clusters. As $t_w$ increases, the finite-size effect comes in the dynamics of $M$. Several clusters begin to merge together as the GCC due to the finite-size effect, which can be treated as temporal percolation. Hence the values of the scaling exponents $x$ and $y$ are conjectured in Sec.~\ref{diffusion}:  
$x=1/\bar{\nu}_p$ and $y=\beta_p/\bar{\nu}_p$, where the critical exponents ($\bar{\nu}_p, \beta_p$) of the percolation universality class in scale-free networks~\cite{Cohen2002} and the extended FSS ansatz with the cutoff exponent $\omega$.

Applying the percolation concept to the dynamics of $M$, we have to look at our numerical data carefully. Assuming $\tau_\times \sim N^{-x}$ with $x=|(3-\gamma_k)/\omega|$ and $M\sim t_w^\eta$ with $\eta=1/(\gamma_k-1)$,  the power-counting analysis of $y$ in Eq.~\eqref{y} is as follows: $y=1/2$ for $\beta=0$ and $y=3/4$ for $\beta=0.5$, which are numerically confirmed in Fig.~\ref{fig5}~(b). Here we note two things: (i) The value of $x$ is different from that in Ref.~\cite{Starnini2014}, due to some modifications of the ADN model. (ii) While for $\beta=0$, the power-counting value of $y$ is exactly same as that of the percolation universality class, for $\beta=0.5$, they are different and numerical data collapse well with the power-counting value. It implies that memory affects dynamic scaling of the GCC. In the static regime, $m\to 1$, so that it is collapsed by the simple scaling form as shown in Fig.~\ref{fig5}~(c). 

To figure out the relative growth of $M$ by the effective size of networks ($N_d$), we also measure $m_d(\equiv M/N_d)$ in the accumulated network up to $t=t_w$. The dynamics of $m_d$ depends on the dynamic topologies of temporal networks. In the highly dynamic regime, $m_d$ decreases as $m_d\sim t_w^{-\eta_d}$ with $\eta_d=1-\eta$ because $N_d \sim t_w$. As time elapses (the time resolution gets wider), we define the extended FSS form of $m_d$ in the intermediate regime, which is similar to Eq.~\eqref{m}:
\begin{align}
\label{m_d}
m_d(t_w,N)\sim N^{-y_d}h(\tau N^{x}),
\end{align}
where $h(X)\sim X^{-\eta_d}$ for $X\ll 1$, so that $y_d=(1-x)\eta_d$ by the power-counting analysis. The results are shown in Figs.~\ref{fig5}(d) and \ref{fig5}(e), where $y_d=1/4$ for $\beta=0$ with $\eta_d=1/3$ and $y_d=1/2$ for $\beta=0.5$ with $\eta_d=2/3$. In the static regime of Fig.~\ref{fig5}(f), $m_d\to 1$ as well. It is noted that the crossover behavior is clearly shown as the minimum in the dynamics of $m_d$. The growth of the GCC is delayed in the presence of memory, which also controls the dynamics of $N_d$. For $\beta=0$, the links mostly contribute to inter-group link formations, while for $\beta=0.5$, the links are used to strengthen the connections within intra-group ones. Fig.~\ref{fig5}(e) show that the relative time of the minimum point gets slower and the finite-size effect comes in later as memory gets stronger.

Using dynamic scaling of the GCC in the intermediate-time regime, we revisit to explain the scaling behaviors of $V$ as shown in Fig.~\ref{fig3}(c) and \ref{fig3}(g). To provide the direct relation between them, we perform the RW process on the time-accumulated network $\widetilde{G}_n=\cup_{i=1}^{i=n}G_i;~n=1, 2, ..., \lceil T/t_w \rceil$. The walker moves in the same way as described in Sec.~\ref{diffusion}. Since $G_n$ is changed to $\widetilde{G}_n$,  $V$ is also changed to $V_g$.   

\begin{figure*}[]
\center
\includegraphics[width=\textwidth]{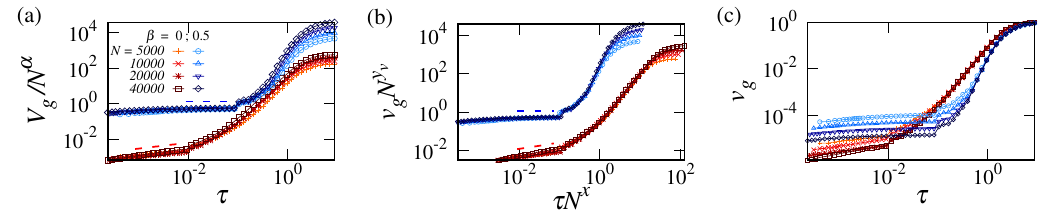}
\caption{We measure the dynamics of $V_g$ in the time-accumulated network, $\widetilde{G}_n;~n=1, 2, ..., \lceil T/t_w \rceil$ at the same setting of Fig.~\ref{fig3}~(c) $t_w=N/100$ for $\beta=0$ and (g) $t_w=N/10$ for $\beta=0.5$: 
(a) We plot $V_g$ as the function of $\tau$, in the context of the same scaling function that suggested in Eq.~\eqref{FSS}, where the slopes of dashed lines represent $\eta_v=1/3$ for $\beta=0$ and $\eta_v=0$ for $\beta=0.5$, respectively. (b) The extended FSS of $v_g(\equiv V_g/N)$ is tested to collapse numerical data near the crossover time, where $y_v=3/4$ for $\beta=0$ and $y_v=1$ for $\beta=0.5$, provided that $x=1/4$ from Figs.~\ref{fig5}(b) and \ref{fig5}(e). Finally, (c) we observe the typical scaling of $v_g$ in the static regime, $v_g\to 1$.}
\label{fig6}
\end{figure*}

Figure~\ref{fig6} represents the scaling behaviors of $V_g$ as the function of $\tau$ and $N$. In Fig.~\ref{fig6}(b), $v_g(\equiv V_g/N)=N^{-y_v}f_g(\tau N^{x})$ with $(x, y_v)=(1/4, 3/4)$ and (1/4, 1) for $\beta=0$ and 0.5, respectively, where $x$ is the same value used in Fig.~\ref{fig5}(b) and \ref{fig5}(e). This corresponds the dynamics of $v_g$ in the intermediate-time regime. The value of $y_v$ can be estimated as $y_v=1+(x-1)\eta_v$ by the power-counting analysis, similar to Eq.~\eqref{y}. In Fig.~\ref{fig6}(a), we reconfirm the results of Fig.~\ref{fig3}(c) and \ref{fig3}(g) in the dynamic regime. In Fig.~\ref{fig6}(c), we check the typical scaling in the static regime.    

Our numerical observation confirms that the temporal percolation concept explains the scaling relation between the dynamics of the GCC and the diffusion of the RW process in temporal networks. So far, we observe that the scaling behaviors depend on both the time resolution and memory, in terms of three different scaling forms. It would be great if there is a single scaling form without any extra tuning, except for the crossover of physical properties. In the next section, we infer the crossover of relevant scaling properties. 

\begin{figure}[b]
\center
\includegraphics[width=\columnwidth]{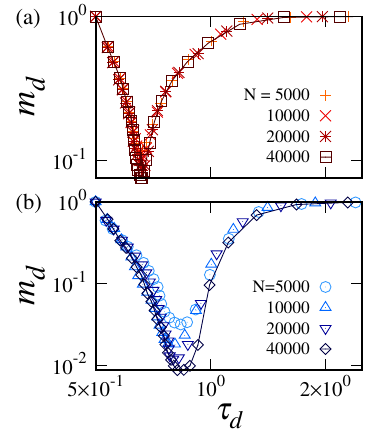}
\caption{The rescaled fraction of the GCC, $m_d(\equiv M/N_d)$, is plotted as the function of the rescaled time $\tau_d(\equiv \tau_w/N_d)$. Without any extra tuning, numerical data collapse well into one curve, which implies that $N_d$ plays a crucial role in the topological change of temporal networks from the dynamic regime to the static regime: (a) $\beta=0$ and (b) $\beta=0.5$. The minimum of the scaling function is located at the $\beta$-dependent value, 
irrespective of $N$. After the rescaled crossover time, $m_d$ monotonically increases up to 1, similar to Figs.~\ref{fig5}(d)--\ref{fig5}(f).}
\label{fig7}
\end{figure}

\section{Fundamental Scaling Properties}
\label{conjecture}

To address the unified scaling behaviors of dynamic topologies in temporal networks, we need to choose fundamental properties, which controls both the effective size of networks  by memory and the time resolution. From the preliminary tests for dynamic scaling of $N_d$ in our early study~\cite{Kim2015}, we find a hint that $N_d$ does depend on both memory and the time resolution. The effective size of the accumulated network is important in determining the crossover from the dynamic regime to the static one. Hence $N_d$ is proper to collapse numerical data in the entire time regime with a minimum point, at which clusters start merging into one big cluster. 

In Fig.~\ref{fig7}, we plot $m_d$ as the function of $\tau_d(\equiv \tau_w/N_d)$ without any extra tuning and numerical data collapse well~\footnote{For $\beta=0.5$, some systematic deviation near and at the minimum point, is due to finite-size corrections to scaling caused by the memory effect.}. It confirms that $N_d$ contains the information of the topological change as time increases and it eventually reaches to the static network size. 

Finally, we propose that one should consider both the size of the GCC ($M$) and the number of distinct nodes ($N_d$) to discuss dynamic scaling in temporal networks. As the time resolution varies, these two quantities reveal key ingredients of dynamic topologies with and without memory as well as the fundamental scaling properties.

\section{Summary and Remarks}
\label{summary}

To summarize, dynamic scaling in temporal networks is proposed with heterogeneous activities and memory, in terms of the modified ADN model. In order to discuss the role of the time resolution $t_w$ and the memory strength $\beta$ in topological changes, we considered the RW process and the percolation picture on time-accumulated networks and conjectured the extended FSS ansatz for dynamic topologies according to $t_w$ and $\beta$. Moreover, we argued scaling exponents in the coverage of the RW process and the dynamics of the largest cluster, namely the GCC. 

As a result, it is found that $t_w$ determines the effective size of networks, while memory controls relevant scaling properties in the vicinity of the crossover from the dynamic regime to the static one. Memory-dependent scaling behaviors are observed in the dynamics of the GCC, the same scaling of which can be applicable to the coverage of the RW process in time-accumulated networks. 
It is implied that the dynamics of the GCC and the effective size of networks are fundamental scaling properties in temporal networks, which play a crucial role in determining dynamic scaling and dynamic topologies. 

Our study is a prototype approach to explain scaling properties in temporal networks. For the better understanding the origin of dynamic scaling according to the time resolution and memory, it is essential to take proper physical quantities and apply the extended FSS theory to them, which allows the systematic analysis. Hence our results would be helpful to those who are interested in scaling properties of dynamic topologies in time-varying systems, such as diffusion and cluster formation. As a possible future study, it would be interesting to speculate dynamic scaling of real-time network data with the burstiness and the periodicity.

\section*{Acknowledgments}
This research was supported by the Basic Science Research Program through National Research Foundation of Korea (NRF)(KR) NRF-2017R1D1A3A03000578 (H.K., M.H.) and NRF-2017R1A2B3006930 (H.K., H.J.).

\appendix
\counterwithin{figure}{section}
\begin{widetext}
\section{Structural Properties of Modified Activity-Driven Network (ADN) Model}
\label{appendix}

The generative process of the modified ADN model, satisfy the following rate equations:
\begin{align}
\frac{ds_i(t)}{dt}&=\frac{a_i}{a_M}+\sum_{j\in\mathcal{N}_i(t)^c}\frac{a_j}{a_M}\left[ s_j(t)^{-\beta}\frac{1}{N-k_j(t)-1}+\delta_{s_j(t),0}\frac{1}{N-1}\right]+\sum_{j\in\mathcal{N}_i(t)}\frac{a_j}{a_M}(1-s_j(t)^{-\beta})\frac{w_{ji}(t)}{\sum_l w_{jl}(t)},
\label{rate-eq-s}\\
\frac{dk_i(t)}{dt}&=\frac{a_i}{a_M}\left[ s_i(t)^{-\beta}+\delta_{s_i(t),0})\right]+\sum_{j\in\mathcal{N}_i(t)^c}\frac{a_j}{a_M}\left[s_j(t)^{-\beta}\frac{1}{N-k_j(t)-1}+\delta_{s_j(t),0}\frac{1}{N-1}\right],
\label{rate-eq-k}
\end{align}
%
where $k_i(t)$ is the accumulated degree of the node $i$ up to time $t$: 
$$k_i(t)= \sum_{j=1}^{N}(1-\delta_{w_{ij}(t),0}),$$ 
where $\delta_{p,q}$ is the Kronecker $\delta$: $\delta_{p,q}=1$ if $p=q$; otherwise, $\delta=0$.
On the right-hand side of Eqs.~\eqref{rate-eq-s} and~\eqref{rate-eq-k}, the first term is for the link generated by the active node $i$, and the rest terms are for the links generated by the other active nodes.

In the asymptotic limit ($1\ll t$ and $k\ll N$), the above equations are considered as
%
\begin{align}
\frac{ds_i(t)}{dt}&\approx \frac{a_i}{a_M}+G_i^{(s)}(t);~ G_i^{(s)}(t)\equiv\sum_{j\in\mathcal{N}_i(t)^c}\frac{a_j}{a_M}s_j(t)^{-\beta}\frac{1}{N}+\sum_{j\in\mathcal{N}_i(t)}\frac{a_j}{a_M}(1-s_j(t)^{-\beta})\frac{w_{ji}(t)}{s_j(t)}, \label{rate-eq-s-2}\\
\frac{dk_i(t)}{dt}&\approx \frac{a_i}{a_M} s_i(t)^{-\beta}+G_i^{(k)}(t);~G_i^{(k)}(t)\equiv\sum_{j\in\mathcal{N}_i(t)^c}\frac{a_j}{a_M}s_j(t)^{-\beta}\frac{1}{N},
\label{rate-eq-k-2}
\end{align}
%
For all $i\in \mathcal{N}=\{1, ...,N\}$, $s_i(t)$ is a monotonically increasing function of $t$, and $G_i^{(s)}(t)$ contains $s_j(t)$ that also has the same equation form of $s_i(t)$.
Hence to consider the leading order term for $t$, Eq.~\eqref{rate-eq-s-2} is expressed as $\frac{ds_i(t)}{dt} \approx \frac{a_i}{a_M} + \mathcal{O}(t^{-\beta}/N)$. For $k\ll N$ limit, we assume $G_i^{(k)}(t)\approx \sum_{j\in\mathcal{N}}\frac{a_j}{a_M}s_j(t)^{-\beta}\frac{1}{N}=\langle \frac{a}{a_M}s(t)^{-\beta}\rangle$ in the Eq.~\eqref{rate-eq-k-2}.
Then we obtain the asymptotic solutions:
\begin{align}
s_i(t)&\sim a_it\nonumber\\
k_i(t)&\sim \left[ a_i^{(1-\beta)}+\langle a^{(1-\beta)}\rangle \right] t^{(1-\beta)},
\label{asym-sol}
\end{align}
where $0\leq \beta <1$.

If the activity distribution is considered as $F(a) \sim a^{-\gamma}$, we find that the distributions of strength and degree in the accumulated network up to $t$, respectively: 
\begin{align}
P(s)\sim s^{-\gamma}~\mbox{and}~
P(k)\sim k^{-\frac{\gamma-\beta}{1-\beta}}.
\label{PsPk}
\end{align}  
As a result, $\gamma_s=\gamma$ for $P(s) \sim s^{-\gamma_s}$ and $\gamma_k=\frac{\gamma-\beta}{1-\beta}$ for $P(k) \sim k^{-\gamma_k}$ where $0\leq \beta<1$. For $\beta=1$: $k_i(t)\sim \ln (a_i t)$ and $P(k)\sim \exp [-c_\gamma(\gamma-1)k]$, where $c_\gamma$ is a positive constant depending on $\gamma$. For the information, it was also reported that the network degree distributions of the ADN model can be changed by memory~\cite{Ubaldi2016}.

Figure~\ref{figA1} represents numerical results for the complementary cumulative distribution functions (CCDFs) of strength, degree, and link-weight for $\gamma=2.5$ and $\beta \in\{0, 0.25, 0.5, 0.75, 1\}$, where $N=T=40000$. As predicted, numerical data are in good agreement with the scaling exponents predicted by the asymptotic solutions.
\begin{figure*}[]
\center
\includegraphics[width=\textwidth]{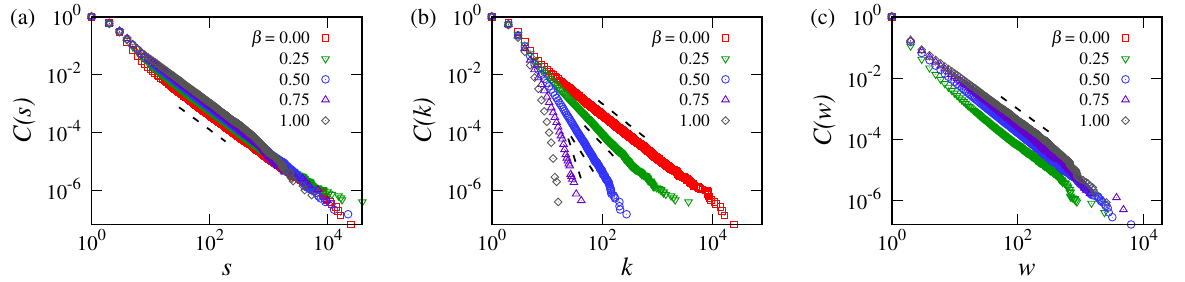}
\caption{The CCDFs of strength, degree, and link-weight are presented in time-aggregated networks with the activity exponent $\gamma=2.5$, where the memory exponent $\beta=0$ ({\color{red} $\square$}), 0.25 ({\color{green} 
$\bigtriangledown$}), 0.5 ({\color{blue} $\bigcirc$}), 0.75 ({\color{violet} $\bigtriangleup$}), and 1 ({\color{gray} $\Diamond$}). (a) The strength CCDF scales as $C(s)\sim s^{-\gamma_s+1}$ with $\gamma_s=2.5$. As expected, they are irrespective of $\beta$. (b) The degree CCDF scale as $C(k)\sim k^{-\gamma_k+1}$. The slopes of dashed lines represent $\gamma_k$ (From top to bottom, $\gamma_k=2.5, 3, 4,$ and 7, respectively). For $\beta=1$, $C(k)$ exponentially decays. (c) The link-weight CCDF is presented as $C(w)$, where the slope of the dashed line is $-1.5$. As long as $\beta$ is not so small, $C(w)$ algebraically decays with the same exponent of $C(s)$. However, for $\beta\ll 1$, $C(w)$ follows no longer a simple power- aw. Eventually, for $\beta=0$, $C(w)=\delta(w-1)$. Numerical data are averaged over 100 network configurations for $N=40000$ and $T=N$.}
\label{figA1}
\end{figure*}

\end{widetext}


%

\bibliography{PRE-RW-ref-final}

%
%
%

\end{document}